\documentclass[sigconf, nonacm]{acmart}

\usepackage{csquotes}
\usepackage{listings}
\usepackage{epigraph}
\usepackage{courier}
\usepackage{xspace}

\newcommand\vldbyear{2024}
\newcommand\vldbworkshop{Relational Models}
\newcommand\vldbauthors{\authors}
\newcommand\vldbtitle{\shorttitle}  
\newcommand\vldbavailabilityurl{https://github.com/BauplanLabs/eudoxia}
\newcommand\vldbpagestyle{plain} 

\newcommand{\Eudoxia}[0]{\textsc{Eudoxia} }
\newcommand{\tinyskip}{\vspace{3pt}}
\newcommand{\mypar}[1]{\tinyskip\noindent\textbf{#1.}\xspace}

\begin{document}
\title{\textit{Eudoxia}: a FaaS scheduling simulator for the composable lakehouse}

\author{Tapan Srivastava}
\authornote{These authors contributed equally: JT led ideation, TS design and implementation. JT, TS, CG all contributed to the final draft.}
\email{tapansriv@uchicago.edu}
\affiliation{%
  \institution{University of Chicago}
  \city{Chicago}
  \state{Illinois}
  \country{USA}
}

\author{Jacopo Tagliabue}
\orcid{0000-0001-8634-6122}
\email{jacopo.tagliabue@bauplanlabs.com}
\authornotemark[1]
\affiliation{%
  \institution{Bauplan Labs}
  \city{New York}
  \country{USA}
}

\author{Ciro Greco}
\email{ciro.greco@bauplanlabs.com}
\affiliation{%
  \institution{Bauplan Labs}
  \city{New York}
  \country{USA}
}

\begin{abstract}

Due to the variety of its target use cases and the large API surface area to cover, a data lakehouse (DLH) is a natural candidate for a composable data system. \textit{Bauplan} is a composable DLH built on ``spare data parts'' and a unified Function-as-a-Service (FaaS) runtime for SQL queries and Python pipelines. While FaaS simplifies both building and using the system, it introduces novel challenges in scheduling and optimization of data workloads. In this work, starting from the programming model of the composable DLH, we characterize the underlying scheduling problem and motivate simulations as an effective tools to iterate on the DLH. We then introduce and release to the community \Eudoxia, a deterministic simulator for scheduling data workloads as cloud functions. We show that \Eudoxia can simulate a wide range of workloads and enables highly customizable user implementations of scheduling algorithms, providing a cheap mechanism for developers to evaluate different scheduling algorithms against their infrastructure.  
\end{abstract}

\maketitle

\pagestyle{\vldbpagestyle}
\begingroup\small\noindent\raggedright\textbf{VLDB Workshop Reference Format:}\\
\vldbauthors. \vldbtitle. VLDB \vldbyear\ Workshop: \vldbworkshop.\\ 
\endgroup
\begingroup
\renewcommand\thefootnote{}\footnote{\noindent
This work is licensed under the Creative Commons BY-NC-ND 4.0 International License. Visit \url{https://creativecommons.org/licenses/by-nc-nd/4.0/} to view a copy of this license. For any use beyond those covered by this license, obtain permission by emailing \href{mailto:info@vldb.org}{info@vldb.org}. Copyright is held by the owner/author(s). Publication rights licensed to the VLDB Endowment. \\
\raggedright Proceedings of the VLDB Endowment. 
ISSN 2150-8097. \\
}\addtocounter{footnote}{-1}\endgroup

\ifdefempty{\vldbavailabilityurl}{}{
\vspace{.3cm}
\begingroup\small\noindent\raggedright\textbf{VLDB Workshop Artifact Availability:}\\
The source code, data, and/or other artifacts have been made available at \url{\vldbavailabilityurl}.
\endgroup
}

\section{Introduction}

\begin{displayquote}
``In Eudoxia, (...), a carpet is preserved in which you can observe the city's true form. At first sight nothing seems to resemble Eudoxia less than the design of that carpet (...), but if you pause and examine it carefully, you become convinced that each place in the carpet corresponds to a place in the city and all the things contained in the city are included in the design.'' (I. Calvino, \textit{Invisible Cities})
\end{displayquote}

The Data Lakehouse (DLH) \cite{Zaharia2021LakehouseAN}, is becoming the \textit{de facto} cloud standard for analytics and Artificial Intelligence (AI) workloads. The DLH promises many improvements over its predecessors, the data lake and warehouse, such as cheap and durable foundation through object storage, compute decoupling, multi-language support, unified table semantics, and governance \cite{mazumdar2023datalakehousedatawarehousing}.

The breadth of DLH use cases makes it a natural target for the philosophy of composable data systems \cite{10.14778/3603581.3603604}. In this spirit, \textit{Bauplan} is a DLH built from ``spare parts'' \cite{Tagliabue2023BuildingAS}: while presenting to users a unified API for assets and compute \cite{10.1145/3650203.3663335}, the system is built from modularized components that reuse existing data tools through novel interfaces: e.g. Arrow fragments for differential caching \cite{10825377}, Kuzu for DAG planning \cite{kuzu}, DuckDB as SQL engine~\cite{duckdb}, Arrow Flight for client-server communication~\cite{arrow_flight}. 

Bauplan serves interactive and batch use cases through a unified Function-as-a-Service (FaaS) runtime running on standard VMs \cite{10.1145/3702634.3702955}. The complexity of resource management in a dynamic, multi-language DLH thus reduces to ``just'' scheduling functions. Building and testing distributed systems is complex, costly, and error-prone in monolithic systems~\cite{daodistributed, whittakerprovenance,liu2008d3s} and is even more so in composable data systems. In order to test our intuitions and safely benchmark policies, we decided to build and release a DLH simulator.

In this work, we present \textsc{Eudoxia}, a scheduling simulator designed for the composable DLH. Our contributions are threefold:

\begin{enumerate}
    \item We describe a composable lakehouse architecture from a programming and execution model perspective, showing how expressing all workloads as functions provides a simple and consistent abstraction for users and the platform alike.
    
    \item We formalize the scheduling problem in this setting and outline the key requirements for any viable solution. 
    
    \item We introduce \Eudoxia as a modular, open-source simulator for this problem space: we detail our design choices, demonstrate typical usage patterns and provide preliminary validation using standard OLAP workloads against cloud production systems.
\end{enumerate}

While \textsc{Eudoxia}'s development was motivated by Bauplan's architecture, we release it to the community\footnote{\url{https://github.com/BauplanLabs/eudoxia}} with a permissive license because we believe its impact to be potentially broader -- either directly as a pluggable module in similar data systems, or indirectly through its abstractions and design principles. 

The paper is organized as follows. In Section \ref{sec:background}, we introduce background on composable DLHs, which serves as the main motivation for this work; Section \ref{sec:design} describes the scheduling problem in detail and presents the high-level structure of the proposed system; Section \ref{sec:101} illustrates how to invoke and run the simulator, how to configure parameters for \Eudoxia, how to register custom scheduling algorithms, and how we validated our approach to have confidence in the results produced by the simulator. We conclude by positioning our work in the context of the existing literature (Section \ref{sec:related}) and of future developments (Section \ref{sec:conclusion}). 

\section{Background and motivation}
\label{sec:background}

The flexibility of serving interactive and batch use cases for both analytics focused (SQL) and AI focused (Python) runtimes is a distinctive feature of DLHs. To motivate the need for a new scheduler simulator, we walk backwards from the developer experience designed to simplify user interaction with heterogeneous workloads (Section~\ref{sec:devex}), and from the corresponding architectural choices (Section~\ref{sec:faas}): as we shall see, the Bauplan DLH is composable and modular at the logical level too.

\subsection{Writing everything as a function}
\label{sec:devex}

In contrast with the hard to learn, difficult to debug Big Data (e.g. Spark
\cite{Wang2021UnderstandingTC,9007378})  and DAG frameworks (e.g. Airflow
\cite{Yasmin2024AnES}),  coding in Bauplan does not require learning new
programming concepts. In particular, data computation can only be expressed
through (SQL or Python) functions with signature \textit{Table(s) -> Table} --
environment variables are either passed as runtime argument (e.g.
\texttt{bauplan run ---namespace xxx}) or stored next to the code
itself.\footnote{To make this work self-contained, we briefly survey here the
relevant DLH pieces. For a fuller background picture on the cloud architecture
and not the developer experience, please see ~\cite{10.1145/3702634.3702955}.} 

We illustrate this with a concrete example. Consider the Bauplan data pipeline comprising following two files, \texttt{parent.sql} and \texttt{children.py}:

\addvspace{\baselineskip}
\begin{lstlisting}[showstringspaces=false,columns=fullflexible,language=SQL,captionpos=b,caption=parent.sql]
-- bauplan_name parent
SELECT col_1, col_2, col_3 FROM raw
\end{lstlisting}
\label{lst:parent}
\addvspace{\baselineskip}

\addvspace{\baselineskip}
\begin{lstlisting}[showstringspaces=false,columns=fullflexible,language=Python,captionpos=b,caption=children.py]
@bauplan.model()
@bauplan.python("3.10", pip={"pandas": "1.5"})
def child(data=bauplan.Model("parent") ):
   return data.do_something()

@bauplan.model(materialize=True)
@bauplan.python("3.11")
def grand_child( data=bauplan.Model("child")):
    return data.do_something()
\end{lstlisting}
\label{lst:children}
\addvspace{\baselineskip}

Pipelines are simply DAGs of functions chained together by naming convention:
the first function to be run is \texttt{parent.sql}, whose input is a
\textit{Table} called \texttt{raw}, which is stored in object storage and
registered in the system catalog. The output of this function is also
represented as a \textit{Table}. This approach is connected to the \textit{dbt}
framework, which pioneered this ``functional'' approach for data analysts
chaining SQL queries together. The second function, \texttt{child}, contained in the
Python file, takes the parent query as input and
produces a new \textit{Table}, which is in turn the input of the final
function. Users interact with data and compute declaratively, using \textit{Bauplan tables} over \textit{data branches}, which are semantic, git-like abstractions over Apache Iceberg tables. As such, the underlying catalog and the data files are abstracted away from users.

We find three major types of interactions between users and DLHs, presented in (roughly) descending order of expected latency:

\begin{enumerate}
    \item \textit{batch data pipelines}: Usually scheduled, these pipelines combine SQL and Python steps and are used in production environments. They prioritize throughput over latency, as no user is actively waiting for results.

    \item \textit{iterative data pipelines}: Triggered during development or debugging, these pipelines benefit from fast feedback loops to improve developer productivity. While not latency-critical in production terms, delays here can slow down iteration speed and increase cognitive load.
    
    \item \textit{interactive queries}: Often issued by analysts or business users in SQL or Python, these queries demand low latency and quick feedback. They represent the “live” interface with data and typically require fast, responsive infrastructure.
\end{enumerate}
 
We see that because functions can read directly from base tables or from the outputs of other functions, each of the above interactions is representable by composing together Python and SQL blocks with specific signatures. 

While the functional abstraction may seem limiting at first, it enables two critical features. First, it lowers the barrier to begin using the system, allowing for example interns with no prior cloud experience to push pipelines to production in their first day of work. Second, within the architecture executing pipelines boils down to orchestrating atomic blocks with the same shape and signature, ``only'' differing by priority.

\subsection{Running everything as a function}
\label{sec:faas}

Users often must use different interfaces to execute each of the different types of DLH workloads (batch, iterative, and interactive) as described in Section~\ref{sec:devex}. 
For example, a user may run a query in a SQL editor supported by a data warehouse, develop in a notebook (supported by a Spark cluster and Jupyter server), and run pipelines as a Spark script on a schedule (supported by a \texttt{submit job} API, a cluster, and an orchestrator). 

\begin{figure}
  \centering
  \includegraphics[width=\linewidth]{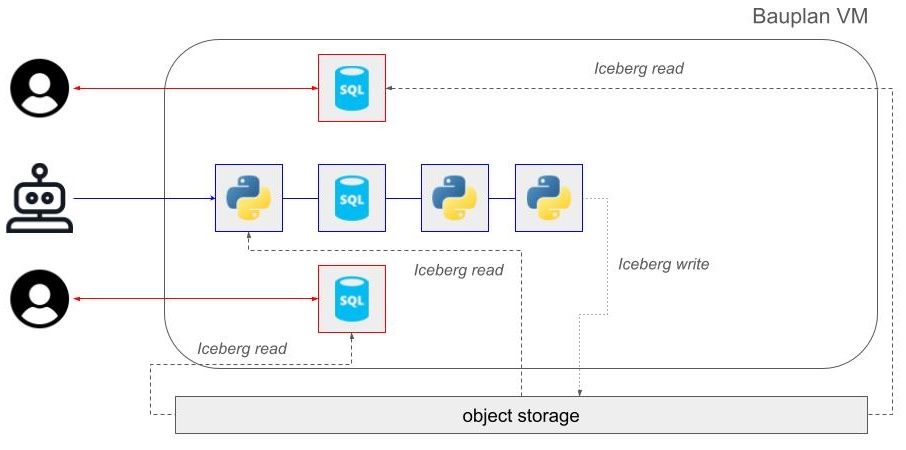}
  \caption{Bauplan workers are off-the-shelf VMs, providing stateless compute capacity over object storage. Within an organization, users and machines (Apache Airflow, AWS Lambda on a schedule etc.) may submit interactive read-only queries (red) or asynchronous read-write pipelines (blue). What scheduling policy for functions can maximize a desired metric (e.g. throughput)?}
  \label{fig:faas}
\end{figure}

Table~\ref{tab:bpln_commands} summarizes the distinctive, composable nature of a
code-first lakehouse. The uniformity at the developer experience level is
mirrored by uniformity at the infrastructure level, where \textit{all}
interactions are served by composing together containerized functions over
object storage as shown in Fig.~\ref{fig:faas}. Due to several optimizations
\cite{10.1145/3702634.3702955}, ephemeral functions spawn in milliseconds inside
off-the-shelf virtual machines (VMs), which greatly simplifies the life-cycle
management of containers. Even system-level actions, such as checking out a data
branch, reading from parquet files to serve a SQL query, or
materializing a result back into the catalog, are written
as functions and are added to the user-specified DAG
by a logical planner \cite{Tagliabue2023BuildingAS}. In other words, any task
executed on Bauplan is a DAG of system-provided and user-specified ephemeral
functions in the view of both the user and the system. No container, warehouse,
or engine exists before or after a request, as any resources or state are spawned on demand.
Indeed, even the additional bidirectional communication required at the end of more interactive workloads are achieved by running an Arrow Flight server as an ephemeral container in the same model as all other functions.

This architecture reframes DLH scheduling as the problem of orchestrating
functions onto pools of resources with varying latency requirements. We find
that the most critical insights we've gained from both intuition and empirical
evidence align with results shared from the systems community: \textit{first},
interleaving interactive and non-interactive workloads end up being more
computational efficient than separating these workloads onto different systems
(i.e. running a query on a warehouse and a pipeline on a Spark cluster)
\cite{10.1145/2741948.2741964,10.1145/3600006.3613155}; \textit{second}, using
functions as building blocks nudges users to write small, re-usable code that is
easier for them to maintain, and importantly easier for the scheduler to reason
about \cite{10.5555/1972457.1972488}. Importantly, existing FaaS schedulers
cannot be re-used \textit{as is} because
these systems--e.g. AWS Lambda~\cite{lambda}, Azure Functions~\cite{azurefunc},
OpenWhisk~\cite{openwhisk}--are designed to support the execution of simple,
fast, stateless, standalone functions with small output sizes.

However, the uniformity of the function interface comes with a trade-off: limited horizontal scaling. While this interface easily supports long pipelines, cross-host communication, and vertical scaling of individual functions \cite{10.1145/3702634.3702955}, each function remains the unit of scheduling and cannot be split across multiple VMs. In traditional big data systems, this has been seen as a limitation, but in practice, many modern workloads can be handled comfortably within a single high-memory VM due to the sharp drop in memory costs (e.g., 1TB fell from \$4K in 2014 to \$1K in 2023~\cite{worlddata}) and the relatively stable size of analytical datasets (i.e. most OLAP workloads today are under 250GB at the \textit{p99.9th} percentile~\cite{Renen-24-Redshift}). This perspective reflects a broader shift toward what some have called “Reasonable Scale” ~\cite{10.1145/3460231.3474604,Tagliabue2023ReasonableSM,McSherry2015ScalabilityBA}, a pragmatic approach that favors simplicity and efficiency over aggressive horizontal scaling at all costs.

In conclusion, we can now see how Bauplan's function-first approach benefits both users and developers. Using system and user functions as the building blocks of the runtime allows a granular understanding of workloads and provides many opportunities to interleave different workload types depending on their latency requirements. Enabling granular scheduling---pausing and resuming DAGs mid-air, moving functions between hosts etc.---is a desired consequence of our architecture but presents a challenge of finding an effective, if not optimal, scheduling algorithm. Considering a DLH that uses a single runtime makes the problem significantly more tractable and simplifies modeling the platform, but there is still a need to evaluate different scheduling algorithms. This motivates our scheduling simulator, \textit{Eudoxia}, which we now discuss. 

\begin{table}[h]
  \begin{tabular}{lll}
    \toprule
    Interaction & UX & Infrastructure\\
    \midrule
    Traditional DLH & &\\
    \midrule
    Batch pipeline & Submit API & One-off cluster\\
    Dev. pipeline & Notebook Session & Dev. cluster\\
    Inter. query & Web Editor (JDBC Driver) & Warehouse\\
    \midrule
    FaaS DLH & & \\
    \midrule
    Batch pipeline & \texttt{bauplan run} & Functions\\
    Dev. pipeline & \texttt{bauplan run} & Functions\\
    Inter. query & \texttt{bauplan query} & Functions\\
  \bottomrule
\end{tabular}
\caption{Interaction types, user interfaces and infrastructure requirements for different DLH designs.}
\label{tab:bpln_commands}
\end{table}

\section{Simulator Design}
\label{sec:design}

We first provide an overview of the motivation and goals behind the simulator in Section~\ref{subsec:design-overview} before discussing the design and major abstractions of the simulator in Section~\ref{subsec:design-abstractions}.

\subsection{Overview}
\label{subsec:design-overview}
A composable lakehouse can be tested in a variety of ways, from cheap but case-based unit and integration tests to very expensive but general formal methods. Simulations rely on a deterministic model of the system (like formal methods) but are low-cost and experimentally driven (like integration tests) and thus are a promising way to evaluate scheduling policies in a complex cloud setup.

Given the FaaS design we adopted (Section~\ref{sec:background}), our scheduling simulator must be able to evaluate different scheduling algorithm implementations both in terms of performance metrics (e.g. throughput and latency) as well as monetary cost (e.g. excess cloud resources or premium storage). A successful solution to this problem will therefore be able to give us confidence over scheduling policies without spending the time and money to evaluate the same policies in a real cloud environment.

\subsection{Design Principles and Major Abstractions}
\label{subsec:design-abstractions}
\begin{figure}
    \centering
    \includegraphics[width=\columnwidth]{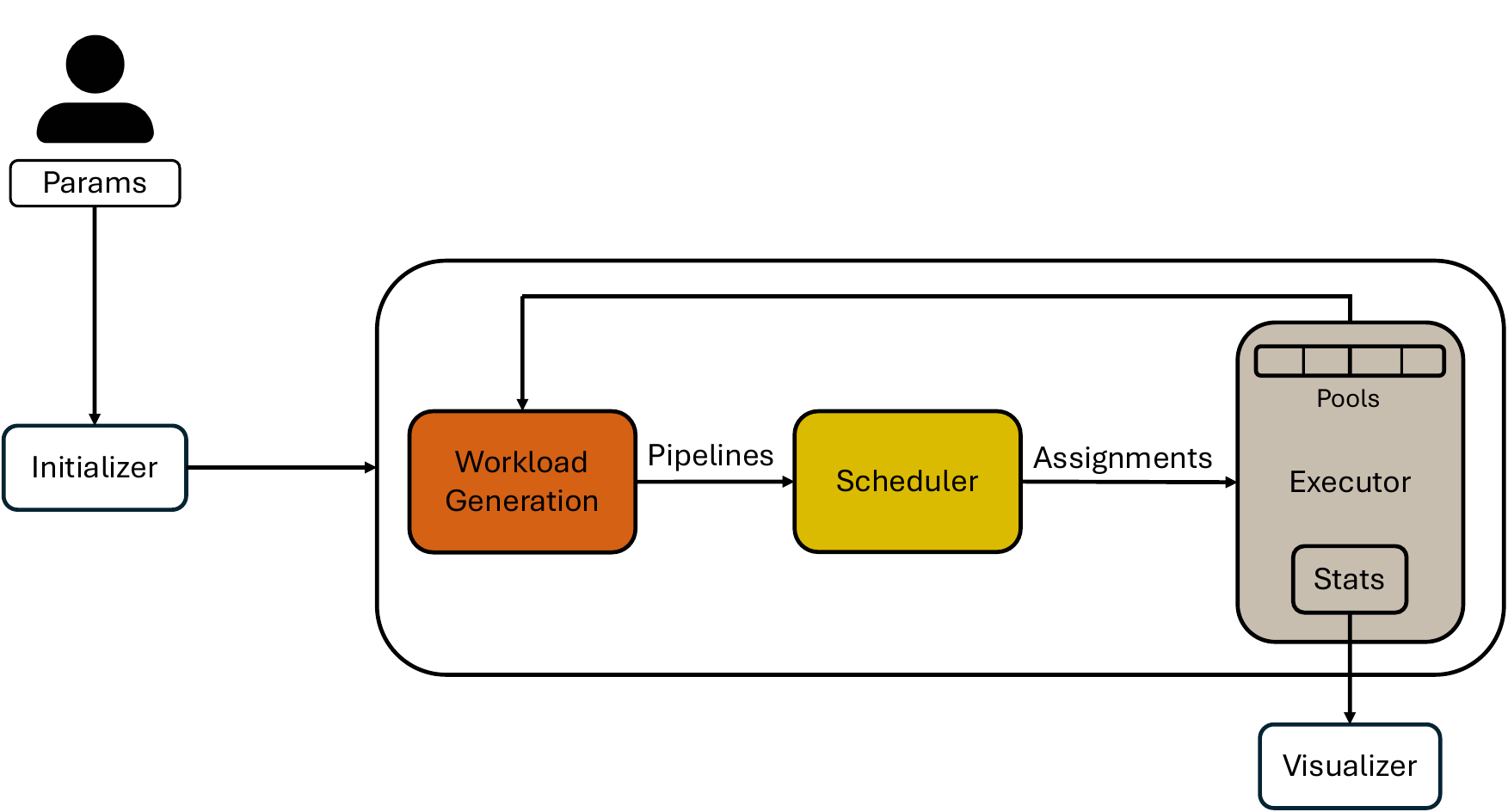}
    \caption{Simulator Architecture. Users set parameters and pass this to the initializer for \Eudoxia which starts a loop of three components, the Workload Generator, Scheduler, and Executor. Once that loop completes, visualizers or other downstream applications can access execution statistics.}
    \label{fig:arch}
\end{figure}

We now present the architecture for our proposed solution. This design, shown in Figure~\ref{fig:arch}, is \emph{modular} to be able to test any scheduling algorithm, to allow for workload customization via parameters set by developers, and to support alternate executor models. 
We decompose this design into three components.
Our simulator operates as a high-level loop, and during each iteration each of these three components complete whatever work is possible for them. Each iteration represents 1 CPU tick or approximately 10 microseconds. 

\subsubsection{Workload Generation}
In a real setup, various users submit pipelines to the system at random intervals.
The workload generator simulates this part of the system by generating pipelines and sending them to the system at user-defined intervals to be scheduled and executed. The workload generator accepts a wide range of parameters which specify how frequently new pipelines arrive, how many resources pipelines require, how long pipelines will take to complete depending on the physical resources (RAM and CPU) allocated to them, among others. Full documentation is available with our artifact. Additionally, this interface allows users to format existing traces and feed them into the simulator rather than generating random ones. 

We model user-submitted pipelines as directed acyclic graphs (DAGs). Each node in a pipeline is called an \emph{operator}, which represents individual functions such as SQL queries or Python functions. Each operator is generated with some required amount of RAM to execute, representing the largest allocation of memory the operator will require to complete. Each operator is also generated with a CPU scaling function, which returns how long the operator will take to complete based on how many CPUs are allocated (for example, a heavy IO task may not scale with CPUs at all, while a stateless filter can scale linearly with more CPUs). Any value associated with a pipeline is randomly drawn from a distribution centered at one of the user-provided (or system default) parameters.
 Finally, each pipeline has one of three Priority Levels, based on the DLH scenarios described in Section~\ref{sec:devex}: in ascending priority order, we have \textit{batch data pipelines}, \textit{iterative data pipelines}, \textit{interactive query}. At each tick when pipelines are generated, they are passed to the scheduler. For most ticks, no new pipelines will be generated.

\subsubsection{Executor}
The user also specifies how many CPUs and RAM are available to allocate to jobs and whether more resources can be accessed for additional monetary cost, i.e. using cloud scaling. The user can specify how many pools of resources there are, what the balance of resources are in each pool, and so on. 

The executor is the manager of these simulated physical resources. We define an abstraction called a \emph{Container} which contains a set of \emph{Operators} to execute and a number of CPUs and amount of RAM. When created each container uses the set of operators provided to calculate how many ticks it will for that container to complete or how many ticks before it will trigger an out-of-memory error based on the parameters in the workload generator and the resources allocated.

\subsubsection{Scheduler}
The Scheduler's responsibility is to allocate resources to sets of Operators (as the Scheduler can subdivide pipelines in allocation) and instruct the Executor on what Containers to create. The Scheduler further has the ability to preempt Containers, instructing the Executor to terminate that container and free up resources. It is the Scheduler's responsibility to manage queues, how allocation decisions are made, what jobs receive allocations sooner than others, and how priority levels are managed. 

Each scheduler implementation must simply match a required type signature: accepting a set of Pipelines from the workload generator, and outputting a list of new Container allocations and Container preemptions to the Executor. At runtime, the user will register different scheduler implementations with the simulator and specify which one it should use during execution.

\section{\Eudoxia 101}
\label{sec:101}

In this section we will describe how users interact with \textsc{Eudoxia}, provide a
sample program and a short description of key parameter options. Then we present a preliminary
validation of our simulator approach using real traces executed on Bauplan.  

\subsection{Developer experience}
We first present how users would start a new \Eudoxia instance (Section~\ref{sec:new-instance}) before presenting the scheduling algorithms already implemented (Section~\ref{sec:scheduler-impls}) and how users can write and register their own implementations (Section~\ref{sec:custom-scheduler}).

\subsubsection{Starting a New Instance}
\label{sec:new-instance}

We aimed to make it as easy as possible to start working with \textsc{Eudoxia}'s
API. To start a simulator instance, users specify input parameters and select a
scheduler implementation, either one of the three scheduler algorithms already
implemented or a custom implementation written in Python and registered at
runtime. Parameters are set in a $TOML$ file, with each parameter in its own
line formatted as \textit{parameter} = \textit{value}. The most important
parameters here the following:\footnote{Full documentation is available with the
code artifact.} 

\begin{itemize}
    \item \textsc{duration}: how many simulated seconds the simulator will run for. Each iteration of the primary loop corresponds to 10 microseconds, intended to roughly approximate the length of 1 CPU cycle. We call each iteration a \textbf{tick}.
    \item \textsc{waiting\_ticks\_mean}: on average how many ticks (10 microseconds) pass between pipelines being generated and sent to the system to be executed.
    \item \textsc{num\_pools}: how many resource pools will exist. In general all available resources are divided evenly among all pools to start.
    \item \textsc{scheduling\_algo}: what scheduling algorithm to use. 
\end{itemize}


Here is how easy is to start an instance: \textsc{run\_simulator} instantiates \Eudoxia with the parameters in \textsc{project.toml}:

\begin{lstlisting}[language=Python, captionpos=b, caption={Minimal code to start a simulation}, label={code:main}]
import eudoxia 
def main():
    paramfile = "project.toml"
    eudoxia.run_simulator(paramfile)
\end{lstlisting}

The \textsc{run\_simulator} method will then begin the core loop described in Section~\ref{sec:design}, containing the workload generation, scheduler, and executor. \Eudoxia will use the \textsc{duration} parameter to compute the number of iterations the loop runs for and will pass each parameter to its appropriate component(s).
 Once \Eudoxia is launched, each component will log its current actions, and CPU and RAM utilization will be logged after each tick for each pool of resources.

\subsubsection{New Scheduling Protocols}
\label{sec:scheduler-impls}
\Eudoxia has three built-in implementations for schedulers.

The first is the \textsc{naive} scheduler, which uses one pool of resources. It assigns all available resources to the next pipeline. When that pipeline completes, it repeats with the next pipeline in the queue.

The next is the \textsc{priority} scheduler, which also assumes one pool of resources. It accounts for both the size of the pool and the priority of pipeline that was submitted (either \textsc{batch}, \textsc{query}, or \textsc{interactive}). New workloads are assigned a container with 10\% of the \emph{total} amount of resources. The scheduler proceeds until it has allocated all resources. If a pipeline completes, those resources are allocated to the next pipeline in the queue. 

If a pipeline fails due to insufficient resources, i.e. an out-of-memory (OOM) error due to insufficient RAM, then those resources are freed but the pipeline re-enters the waiting queue of the scheduler with information about what resources were allocated to the container which failed. 

If a previously-failed pipeline arrives, the scheduler attempts to double the resources previously allocated up to a maximum of 50\% of total CPU or RAM, at which point the scheduler returns the failure to the user. If there are not sufficient available resources to double the allocation, the job is put back on the queue to wait. 

Finally, in the event that all resources are allocated and a high priority pipeline, such as a \textsc{query}, arrives, the scheduler scans the currently running containers for any which are running a low-priority job (such as a \textsc{batch} workload). That container is preempted, freeing its resources to be used for the \textsc{query}. The \textsc{batch} pipeline is put back on the waiting queue with a log of what resources were last allocated to it; however, this pipeline does not also receive the flag indicating it failed. So when the \textsc{batch} pipeline next arrives the scheduler will allocate the \emph{same} resources it allocated previously.

The third scheduling algorithm is the \textsc{priority-pool} scheduler. This operates similarly to the \textsc{priority} scheduler but with multiple resource pools in the Executor. Every time the Scheduler considers a new pipeline, it identifies which pool has the most available resources and allocates a container on that pool. It also handles preemption in the same way, but this time on multiple pools. 


\subsubsection{Registering New Scheduler Implementations}
\label{sec:custom-scheduler}
\Eudoxia allows users to write custom scheduler implementations by following three simple steps: writing an initialization function, writing a scheduler function, and using two decorators.


The \textit{initialization function} accepts one parameter, an instance of the \textsc{Scheduler} class, and returns nothing. This function initializes any needed data structures within the \textsc{Scheduler}. 

The \textit{scheduler function} \emph{must} accept three parameters and return two values. The three parameters are:

\begin{enumerate}
    \item An instance of the \textsc{Scheduler} class. 
    \item A list of pipelines which failed in the previous tick
    \item A list of pipelines which were newly created in this tick
\end{enumerate}

In general, the list of newly created pipelines is often empty, as the workload generation step creates new pipelines at random intervals. Similarly, the list of failures only includes jobs which the executor failed, such as for an out-of-memory error. This does not include pipelines which the scheduler preempted. If the scheduler wishes to preempt pipelines it must manage those queues itself to ensure no pipelines fall through the cracks.

Finally, the algorithm must return two values:

\begin{enumerate}
    \item \textsc{Suspensions}: these are a set of pipelines that the scheduler is instructing the Executor to preempt so that its resources may be freed. For the priority scheduler, it places pipelines to be preempted in an internal \textsc{suspending} queue, which after one tick it moves back into the standard waiting queues. 
    \item \textsc{Assignments}: the second return value is a list of new assignments instructing the Executor what resources to allocate to a container and what job to run inside that container.    
\end{enumerate}

Putting these requirements together, extending \Eudoxia with a custom scheduler is as simple as the snippets below -- note how the \textit{two decorators} in \textsc{algorithm.py} and the parameter in \textsc{project.toml} reference the same \textit{key}:\footnote{Users should import \textsc{scheduler\_init} and \textsc{scheduler\_algo} in \textsc{main.py} first so that the decorators register the keys before the instance starts.} 

\begin{lstlisting}[language=Python, captionpos=b, caption={algorithm.py: \textit{scheduler function}.}, label={code:algo}]

from eudoxia.core import Scheduler
from eudoxia.core import Failure, Assignment, Pipeline
from eudoxia.algorithm import register_scheduler, register_scheduler_init
from typing import List

@register_scheduler_init(key="my-scheduler")
def scheduler_init(sch: Scheduler): 
    ...

@register_scheduler(key="my-scheduler")
def scheduler_algo(sch: Scheduler, f: List[Failure], p: List[Pipeline]): 
    ...
    return suspends, assignments
    
\end{lstlisting}

\begin{lstlisting}[language=Python, captionpos=b, caption={project.toml: custom parameter.}, label={code:toml}]
scheduling_algo = "my-scheduler"
\end{lstlisting}

\begin{lstlisting}[language=Python, captionpos=b, caption={main.py: custom imports and instantiation.}, label={code:full_main}]

from algorithm import scheduler_init, scheduler_algo
import eudoxia
def main():
    paramfile = "project.toml"
    eudoxia.run_simulator(paramfile)
\end{lstlisting} 

\subsection{Preliminary Validation}

While developer experience, clarity in abstractions and extensibility are
crucial aspects for its adoption, \textsc{Eudoxia's} utility depends on
reliability and robustness. 

The simulation generates pipelines which
have two key values: (1) how the number of CPUs allocated impacts the pipeline's execution time (if at all) 
and (2) the minimum RAM allocation needed to avoid an out-of-memory error. 
The scheduling algorithms do not have access to these values or scaling functions; however, once the pipeline is allocated to a container of resources, those
values are used to determine what the true execution time for a pipeline on a container will be. We believe that this is a realistic setup that can effectively represent any
kind of workload in the appropriate and necessary dimensions. 

\begin{figure}
    \centering
    \includegraphics[width=\columnwidth]{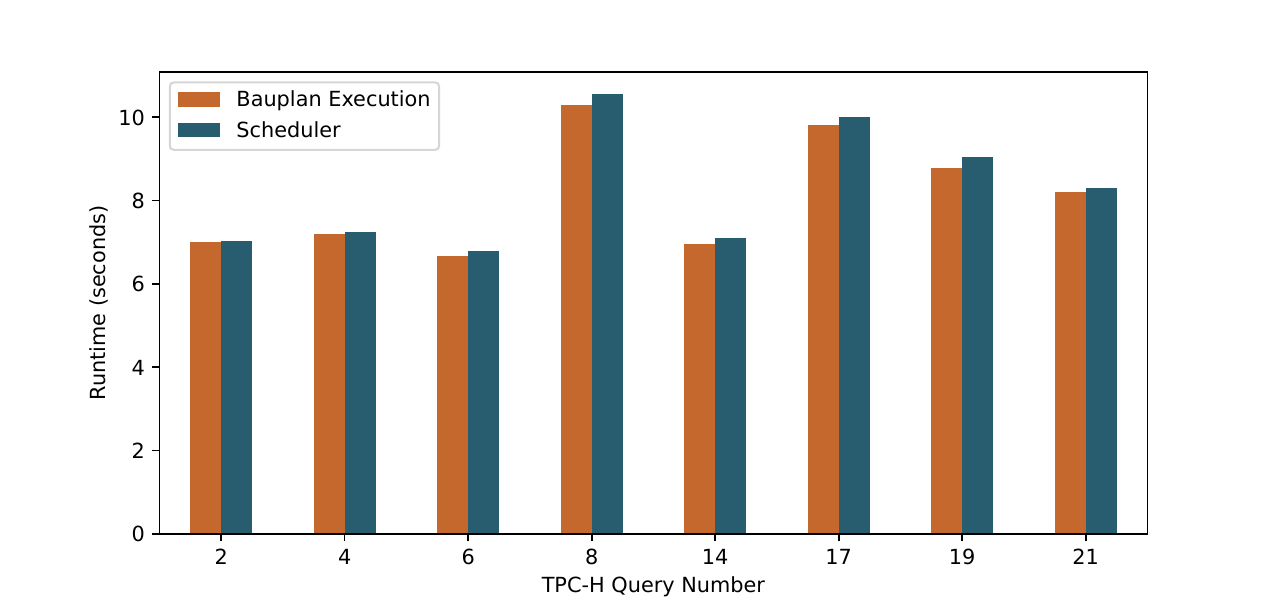}
    \caption{Distribution of percent error of simulator estimates for runtime vs. real runtime for TPC-H queries at 10GB.}
    \label{fig:boxplot}
\end{figure}

We first validate this approach by running data workloads against a Bauplan
cloud instance, measuring runtime statistics such as CPU and RAM utilization along
with runtime and comparing this to the runtime estimated by \Eudoxia on a
pipeline with similar statistics. We run the
common data analytics benchmark TPC-H~\cite{tpch} and run its 22 queries against a 10GB dataset on a instance running on an AWS c5ad.4xlarge instance with 16 vCPUs and 32GB of RAM. As described, each query is compiled by Bauplan into a small number of execution blocks (i.e. functions), and we observe CPU and RAM usage during execution. Each query is run
alone on the instance, and we disable caching. For three queries (11, 16, and
22), the runtime was so short that resource utilization statistics could not be
gathered from underlying telemetry systems. The percent error in runtime of the
scheduler versus the true runtime as executed on a Bauplan instance ranges from
0.44\% to 3.08\% with an average of 1.74\% error. We additionally plot the real
and simulated runtimes for a subset of TPC-H queries for ease of visual
interpretation in Figure~\ref{fig:boxplot}. We see that the simulated runtime well approximates the true execution time, indicating that the simulator can be relied upon to give realistic results.  

Furthermore, because \Eudoxia supports varying CPU scaling functions and enables real traces to be plugged in rather than using random generation, the system can easily emulate how the benchmark's performance would vary if different compute resources are allocated or if the benchmark ran on larger datasets. The modular design enables users to reproduce other results cheaply, test how algorithms would hold up against different types of workloads, or consider how a current implementation would fare against a changing setup. 

\section{Related Work}
\label{sec:related}

\mypar{Composable data systems} The FaaS lakehouse modeled by \Eudoxia is built in the \textit{composable data system} tradition \cite{10.14778/3603581.3603604}. In a sense, the deconstructed lakehouse \cite{Tagliabue2023BuildingAS} is the natural generalization of the ``Deconstructed Database'' \cite{khurana2018modern}. The rapid growth of \textit{DataFusion} \cite{10.1145/3626246.3653368} in the composable data community is fostering an eco-system of novel single-node systems \cite{InfluxData,Arroyo} that could benefit from the simulation methodology and code in \Eudoxia.

\mypar{Cloud Scheduling}
There is a broad range of work in the realm of scheduling or scheduling workloads in cloud environments that is relevant to \Eudoxia. Motlagh et. al.~\cite{curinoworkload} provides an analytical framework to evaluate different scheduling approaches. Similarly, Hai et. all~\cite{hai2023task} proposes a new scheduling approach but does so with a broad cloud usage pattern in mind. In contrast, \Eudoxia is designed for specifically a data lakehouse/composable data system environment. Rather than trying to survey a range of approaches or techniques, \Eudoxia focuses specifically on an application deployment on a single Bauplan instance on an EC2 node. 

Another common goal for schedulers is to abide by quality-of-service (QoS) guidelines. While this is a vital part of the cloud ecosystem, our goal behind \Eudoxia was to consider, experiment, and use that simulator to evaluate future scheduling approaches. 

Finally, a broad range of literature covers scheduling under power constraints. Specifically, as power-constrained applications throttle query performance in some instances as they limit CPU frequency, etc. There is a broad range of work in this area, including \cite{linpower, cloudsim, rountree, bailey, rountree_hpc}. However, \Eudoxia is generally uninterested in how power consumption limits resources availability and workload runtime, as blob storage and VM services offered by cloud vendors abstract away the power consumption needs for cloud infrastructure. 

\section{Conclusion}
\label{sec:conclusion}

In this paper, we described a composable lakehouse, \textit{Bauplan}, from the perspective of its programming and execution model, which reduced the main use cases to scheduling functions with different priorities. While the developer experience gets simplified, we face greater optimization challenges compared to general purpose FaaS systems due to our target workload, which prevented us from re-using existing FaaS schedulers. As simulations are a cost-effective ways to test cloud distributed systems, we introduced our simulator, \Eudoxia, to enable robust but cheap evaluation of the effect of different scheduling algorithms. Through concrete examples, we described \Eudoxia design principles and developer experience, and we provided a preliminary quantitative validation running standard OLAP benchmarks in production and in the simulator. 

Eudoxia is easy to extend and applicable beyond its initial
simulation scope. For example, plugging real-world scaling functions estimated
from traces is trivial and may be useful for other use cases. In the same vein,
while most of our simulations are single pool because of Bauplan's design,
benchmarking functions across capacity pools is a useful feature in
distributed systems. We release the code to the community with a permissive
license as we believe our abstractions and lessons -- if not \Eudoxia itself -- to
be of use to the broader composable data system community. 

\clearpage

\bibliographystyle{ACM-Reference-Format}
\bibliography{sample}

\end{document}